\documentclass[11pt]{article}   
\usepackage{amssymb,latexsym}
\usepackage{epsfig}
\usepackage{eufrak}
\usepackage{amsmath}
\usepackage{mathrsfs}
\usepackage{color}

\newtheorem{theorem}{Theorem}
\newtheorem{lemma}{Lemma}
\newtheorem{corollary}{Corollary}

\newcommand{\be}{\begin{equation}}
\newcommand{\ee}{\end{equation}}
\newcommand{\bee}{\begin{eqnarray*}}
\newcommand{\eee}{\end{eqnarray*}}
\newcommand{\bel}{\begin{eqnarray}}
\newcommand{\eel}{\end{eqnarray}}
\newcommand{\bec}{\begin{cases}}
\newcommand{\eec}{\end{cases}}
\newcommand{\bem}{\begin{bmatrix}}
\newcommand{\eem}{\end{bmatrix}}

\newcommand{\la}{\label}
\newcommand{\li}{\left}
\newcommand{\ri}{\right}

\newcommand{\udl}{\underline}

\newcommand{\lf}{\lfloor}
\newcommand{\rf}{\rfloor}

\newcommand{\de}{\delta}

\newcommand{\ze}{\zeta}
\newcommand{\al}{\alpha}
\newcommand{\ba}{\beta}

\newcommand{\f}{\frac}

\newcommand{\cd}{\cdots}

\newcommand{\qu}{\quad}
\newcommand{\qqu}{\qquad}

\newcommand{\mscr}{\mathscr}

\newcommand{\bb}{\mathbb}

\newcommand{\wh}{\widehat}

\newcommand{\mrm}{\mathrm}

\newcommand{\tx}{\text}

\newcommand{\iy}{\infty}

\newcommand{\bed}{\begin{description}}
\newcommand{\eed}{\end{description}}
\newcommand{\bei}{\begin{itemize}}
\newcommand{\eei}{\end{itemize}}
\newcommand{\ben}{\begin{enumerate}}
\newcommand{\een}{\end{enumerate}}
\newcommand{\bib}{\bibitem}
\newcommand{\beL}{\begin{lemma}}
\newcommand{\eeL}{\end{lemma}}
\newcommand{\beT}{\begin{theorem}}
\newcommand{\eeT}{\end{theorem}}
\newcommand{\beC}{\begin{corollary}}
\newcommand{\eeC}{\end{corollary}}
\newcommand{\sect}{\section}

\newcommand{\bpf}{\begin{pf}}
\newcommand{\epf}{\end{pf}}

\newcommand{\pfbox}{\hfill\mbox{$\Box$}}

\newenvironment{pf}{\paragraph*{Proof{\rm.}}}{\pfbox\bigskip}

\begin{document}

\title{{\bf New Sequential Methods for Detecting Portscanners}
\thanks{The author had been previously working with
Louisiana State University at Baton Rouge, LA 70803, USA, and is now with Department of Electrical Engineering, Southern University and A\&M
College, Baton Rouge, LA 70813, USA; Email: chenxinjia@gmail.com.} }

\author{Xinjia Chen}

\date{First submitted in April 2012}

\maketitle

\begin{abstract}

In this paper, we propose new sequential methods for detecting port-scan attackers which routinely perform random ``portscans'' of IP addresses
to find vulnerable servers to compromise.  In addition to rigorously control the probability of falsely implicating benign remote hosts as
malicious, our method performs significantly faster than other current solutions.  Moreover, our method guarantees that the maximum amount of
observational time is bounded.  In contrast to the previous most effective method, Threshold Random Walk Algorithm,  which is explicit and
analytical in nature, our proposed algorithm involve parameters to be determined by numerical methods.  We have developed computational
techniques such as iterative minimax optimization for quick determination of the parameters of the new detection algorithm.  A framework of
multi-valued decision for testing portscanners is also proposed.

\end{abstract}

\sect{Introduction}

As Internet becomes pervasive to our society,  it is increasingly important to  develop high performance network intrusion detection system
(NIDS) to identify an attacker  to allow for protective response to mitigate or fully prevent damage. An important need in such NIDS is prompt
response: the sooner a NIDS detects malice, the lower the resulting damage. At the same time, a NIDS should not falsely implicate benign remote
hosts as malicious \cite{Heberlein, Jung, Leckie}.  There are many types of network intrusions.  An extremely dangerous one is the ``portscans''
intrusion.  A port-scan is an attack that sends client requests to a range of server port addresses on a host, with the goal of finding an
active port and exploiting a known vulnerability of that service \cite{Roesch, Staniford, Yegneswaran}.

In recent years, some detection schemes have been developed by virtue of statistical hypothesis testing.  For example, the problem of detecting
port-scan attacks has been addressed in the framework of testing a binomial parameter. In this direction, adaptive methods such as the
Sequential Probability Ratio Tests \cite{Wald} have been explored for fast detection  of  port-scan attacks. However, these techniques generally
suffers from two drawbacks. First, the maximum number of required observations is not deterministically bounded.  Hence, there is a probability
that the detection time is extremely long.  Second, the existing detection algorithms usually attempt to be optimal for only a few parametric
values and consequently the average performance for other parametric values many be very poor.   In order to overcome these limitations, we
propose a new methods for fast detection of port-scan attacks in the general framework of multistage tests of hypotheses.

The remainder of the paper is organized as follows.  In Section 2, we consider the problem of testing port-scan attack.  In particular, we
discuss  the widely accepted binomial model and the threshold random walk detection algorithm.  In Section 3, we introduce new sequential
algorithm for detecting port-scan attacks. In Section 4, a framework of multi-valued decision for testing portscanners is proposed. Section 5 is
the conclusion.

\section{Binomial Model}

A major characteristics of scanners is that they have higher chance than legitimate remote hosts to choose hosts which do not exist or do not
have the requested service activated, since they lack precise knowledge of which hosts and ports on the target network are currently active
\cite{Jung, Leckie, Yegneswaran}. Based on this observation, a detection problem has been formulated to provide the basis for an on-line
algorithm whose goal is to reduce the number of observed connection attempts (compared to previous approaches) to flag malicious activity, while
bounding the probabilities of missed detection and false detection. In this direction, a widely accepted model is the binomial model \cite{Jung,
Staniford} described in the sequel.

We shall adopt the description of \cite{Jung} for the binomial model used for the detection of port-scan attacks.  The activity that a remote
source $r$ makes a connection attempt to a local destination $l$ can be considered as a random event.  A frequent method to model such event is
to classify the outcome of the attempt as either a ``success'' or a ``failure'', where the latter corresponds to a connection attempt to an
inactive host or to an inactive service on an otherwise active host.  More formally,  for a given $r$, let $X_i$ be a random variable that
represents the outcome of the first connection attempt by $r$ to the $i$-th distinct local host, where \be \la{model}
X_i = \bec 1 & \tx{if the connection attempt is a success},\\
0  & \tx{if the connection attempt is a failure} \eec \ee As illustrated in \cite{Jung, JungB}, it is reasonable to assume that $X_i, \; i =
1,2, \cd$ are independent and identically Bernoulli random variables such that
\[
 \Pr \{ X_i = 1 \} = 1 - \Pr \{ X_i = 0 \} = p, \]
 where $p \in (0, 1)$ is the success rate of making a connection. Usually, the success rate $p$ is
 unknown and varying for different types of users.  However, the success rate $p$ of a scanner is normally very low, while the success rate $p$ of a
 benign user is high.  By appropriate choosing values of threshold values $p_0$ and $p_1$ such that $0 < p_0 < p_1 < 1$ based on
empirical data analysis of relevant networks, the hypothesis that ``the host is a scanner'' can be formulated as $\mscr{H}_0: p \leq p_0$.
Similarly, the hypothesis that ``the host is a benign user'' can be formulated as $\mscr{H}_1: p \geq p_1$.  This amounts to the problem of
testing statistical hypotheses
\[
\mscr{H}_0: p \leq p_0 \qu \tx{versus} \qu  \mscr{H}_1: p \geq p_1 \] based on $X_i, \; i = 1,2, \cd$.  Throughout the remainder of this paper,
let $\Pr \{ E \mid p \}$ denote the probability of event $E$ associated with $p$.  To control the probabilities of making wrong decisions, it is
typically required that \be \la{riskspec}
 \Pr \{ \tx{Reject} \; \mscr{H}_0 \mid p \} \leq \al \; \tx{for all} \; p \in (0, p_0], \qu
  \Pr \{ \tx{Reject} \; \mscr{H}_1 \mid p \} \leq \ba \; \tx{for all} \; p \in [p_1, 1)
\ee where $\al, \ba \in (0, 1)$ are some pre-specified numbers. In order to minimize the potential damage of network intrusion and control the
probability of false alarm, it is desirable to make this detection as quickly as possible, but with a high probability of being correct.  The
above formulation of the port-scanner detection problem has been proposed by a number of researchers and many detection algorithms have been
developed.  One of the most effective algorithms for early scan detection is the Threshold Random Walk Algorithm (TRWA) developed in \cite{Jung,
JungB}, which is represented in the following section.

\section{Threshold Random Walk Algorithm}

The widely cited Threshold Random Walk Algorithm \cite{Jung} is derived from the famous Sequential Probability Ratio Test (SPRTs) invented by
Abraham Wald \cite{Wald} in the War time in response to the demand of efficient testing of ammunition power.  Define relative frequency
$\wh{p}_n = \f{ \sum_{i=1}^n X_i }{n}$ for $n = 1, 2, \cd$. The idea of TRWA is to continuously observe the probability ratio {\small
\[
\f{ \Pr \{ X_1, \cd, X_n \mid p_0 \} }{ \Pr \{ X_1, \cd, X_n \mid p_1 \}  } = \exp \li ( n \li [ \wh{p}_n \ln \f{p_0}{p_1} + (1 - \wh{p}_n) \ln
\f{1 - p_0}{1 - p_1} \ri ] \ri ) \] } for $n = 1, 2, \cd$. The observational process is continued until {\small $\f{ \Pr \{ X_1, \cd, X_n \mid
p_0 \} }{ \Pr \{ X_1, \cd, X_n \mid p_1 \}  } \leq k_0$} or {\small $\f{ \Pr \{ X_1, \cd, X_n \mid p_0 \} }{ \Pr \{ X_1, \cd, X_n \mid p_1 \}  }
\geq k_1$} for some positive integer $n$, where $k_0 < k_1$ are two pre-specified positive integers for controlling the probability of making
wrong decisions. At the termination of the observational process, a decision is made as follows:

If {\small $\f{ \Pr \{ X_1, \cd, X_n \mid p_0 \} }{ \Pr \{ X_1, \cd, X_n \mid p_1 \}  } \leq k_0$}, then declare the source $r$ as a benign
user. If {\small $\f{ \Pr \{ X_1, \cd, X_n \mid p_0 \} }{ \Pr \{ X_1, \cd, X_n \mid p_1 \}  } \geq k_1$},  then declare the source $r$ as a
scanner.

It can be shown that TRWA has the following properties:  If $0 < k_0 = \al < 1 < \f{1}{\ba} = k_1$, then the TRWA ensures the risk requirement
(\ref{riskspec}). Moreover, the average number of observations is minimized for both $p_0$ and $p_1$ among all possible tests such that $\Pr \{
\tx{Reject} \; \mscr{H}_0 \mid p_0 \} \leq \al$ and $\Pr \{ \tx{Reject} \; \mscr{H}_1 \mid p_1 \} \leq \ba$.

Despite its remarkable simplicity and optimality for threshold values,  the TRWA has the following major drawbacks. First, the number of
observations is not bounded by a deterministic number. In the extreme case, the detection time can be unacceptably long. Second,
 as a consequence of the fact that TRWA is optimal when the true success rate $p$ assumes value $p_0$ or
$p_1$,  the average performance can be very poor when the true rate of success differs from $p_0$ and $p_1$.  Since the choice of threshold
values $p_0$ and $p_1$ is based on empirical data analysis and  is thus some what arbitrary, the performance of the detection algorithm is
important for $p$ taking values different from $p_0$ and $p_1$.  To overcome these drawbacks, we propose to develop a detection method in the
next section.

\sect{ New Detection Algorithm}

Our new detection algorithm depends on $3$ positive parameters $a, b$ and $\ze$,  which are to be determined by a computational method to
guarantee the risk requirement.  The parameter $\ze$ is called the {\it risk tuning parameter}.  The parameters $a$ and $b$ are referred to as
{\it weighting coefficients}.  Let the relative frequency $\wh{p}_n$ be defined as before.  For the ease of describing our detection algorithm,
define new random variables
\[
Y_n = \bec \ln \f{1}{1 - p_0} & \tx{for} \; \wh{p}_n = 0,\\
\wh{p}_n \ln \f{ \wh{p}_n }{p_0} + (1 - \wh{p}_n) \ln \f{1 - \wh{p}_n }{1 - p_0} & \tx{for} \; 0 < \wh{p}_n < 1,\\
\ln \f{1}{p_0} & \tx{for} \; \wh{p}_n = 1 \eec \]
\[
Z_n = \bec \ln \f{1}{1 - p_1} & \tx{for} \; \wh{p}_n = 0,\\
\wh{p}_n \ln \f{ \wh{p}_n }{p_1} + (1 - \wh{p}_n) \ln \f{1 - \wh{p}_n }{1 - p_1} & \tx{for} \; 0 < \wh{p}_n < 1,\\
\ln \f{1}{p_1} & \tx{for} \; \wh{p}_n = 1 \eec \] for $n = 1, 2, \cd$. We are now in a position to state the stopping and decision rules of our
detection algorithm in the sequel.

\subsection{Stopping and Decision Rules} \la{strule}

Assume that the risk tuning parameter and weighting coefficients can be determined to satisfy the risk requirement (\ref{riskspec}), our
detection algorithm can be described as follow.

{\it Continue taking observations until {\small $Y_n \geq \f{1}{n} \ln \f{1}{\ze a}, \; \wh{p}_n \geq p_0$} or {\small $Z_n \geq \f{1}{n}  \ln
\f{1}{\ze b}, \; \wh{p}_n \leq p_1$} for some positive integer $n$.  At the termination of observational process, make the following decision:
If {\small $Z_n \geq \f{1}{n}  \ln \f{1}{\ze b}, \; \wh{p}_n \leq p_1$}, then declare the source $r$ as a scanner. If {\small $Y_n \geq \f{1}{n}
 \ln \f{1}{\ze a}, \; \wh{p}_n \geq p_0$}, then declare the source $r$ as a benign user}.

\begin{figure}[here]
\centering
\includegraphics[height=8cm]{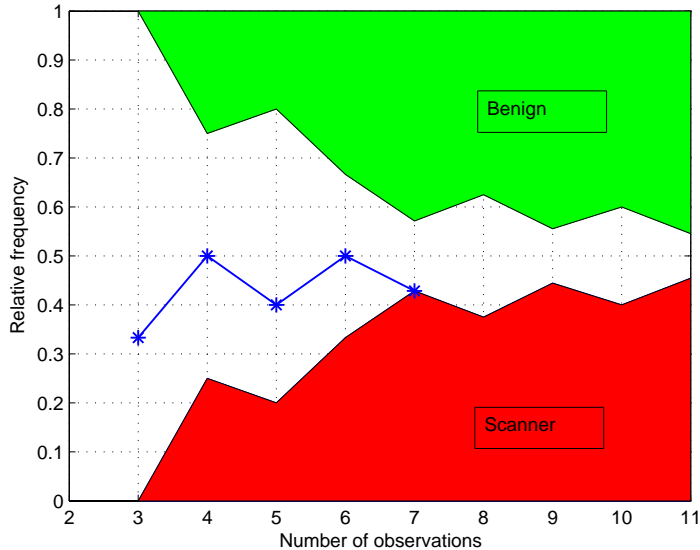}
\caption {An illustration of new detection algorithm}
\label{fig_Detection}       
\end{figure}

For $p_0 = 0.2, \; p_1 = 0.8$, our stopping and decision rules with $\ze = 1$ and $a = b = 0.1$ can be shown by Figure \ref{fig_Detection}. The
lower shaded area represents the acceptance region of $\mscr{H}_0$. The upper shaded area represents the rejection region of $\mscr{H}_0$. The
blue line with star symbols represents a sample path.  The observational process is continued until the sample path hit either the acceptance
region of rejection region of $\mscr{H}_0$.  If the sample path hits the acceptance region of $\mscr{H}_0$, then declare that $r$ is a scanner.
If the sample path hits the rejection region of $\mscr{H}_0$, then declare that $r$ is a benign user.

\subsection{Determination of Risk Tuning Parameter and Weighting Coefficients}

Given that our detection algorithm can be parameterized as in Section \ref{strule}, we need to determine the risk tuning parameter $\ze$ and
weighting coefficients $a, b$ so that the required number of observations is as small as possible, while guaranteeing the risk requirement
(\ref{riskspec}). The computational process for accomplishing this task is called {\it risk tuning}.  Clearly, the risk requirement is satisfied
if $\ze$ is sufficiently small. This implies that if the weighting coefficients are given, one can determine the risk tuning parameter $\ze$ to
meet the risk requirement by the following two steps: First, find the maximum number, $\udl{\ze}$, in the set $\{ 2^{-i}: i \in \bb{N} \}$,
where $\bb{N}$ is the set of natural numbers,  such that the risk requirement  is satisfied when the risk tuning parameter $\ze$ assumes value
$\udl{\ze}$. Second, apply a bisection search method to obtain a number $\ze^\star$ as large as possible from interval $[\udl{\ze}, 2
\udl{\ze})$  such that the risk requirement is satisfied when the risk tuning parameter $\ze$ assumes value $\ze^\star$.  However, these two
steps are not sufficient to produces detection algorithm of satisfactory efficiency if the weighting coefficients are not properly chosen.   To
overcome this limitation, we observe that to make a detection algorithm efficient, it is an effective approach to make the detection algorithm
efficient when the success rate $p$ assumes values $p_0$ and $p_1$. This is a consequence of the fact that $\Pr \{ \tx{Accept} \; \mscr{H}_0
\mid p \}$ is non-increasing with respect to $p \in (0, 1)$. Due to the monotonicity of the operating characteristic function, it suffices to
ensure $\Pr \{ \tx{Reject} \; \mscr{H}_0 \mid p_0 \} \leq \al$ and $\Pr \{ \tx{Reject} \; \mscr{H}_1 \mid p_1 \} \leq \ba$ to satisfy the risk
requirement (\ref{riskspec}).  Define \[ A = \f{ \al } { \Pr \{ \tx{Reject} \; \mscr{H}_0 \mid p_0 \} }, \qqu B = \f{ \ba } { \Pr \{ \tx{Reject}
\; \mscr{H}_1 \mid p_1 \} }
\]
\[
Q = \max \li \{ A, \; B \ri \}, \qqu R = \min \li \{ A, \; B \ri \} \]
 as functions of $a, \; b$ and $\ze$. For purpose of developing an
efficient detection algorithm satisfying the risk requirement, we propose to determine risk tuning parameter $\ze$ and weighting coefficients
$a, \; b$ such that $Q$ is minimized under the constraint that $R$ is no less than $1$. This task can be accomplished by applying the iterative
minimax optimization algorithm described as follows.

\vspace{0.05in}

\begin{tabular} {|l |}
\hline

$\nabla  \; \tx{Set the maximum number of iterations as $k_{\mrm max}$.    Choose the initial values of}$\\
$ \qu \tx{weighting coefficients as $a = \al$ and $b =
\ba$. Let $\wh{Q} \leftarrow \iy$ and $k \leftarrow 0$}$.\\
$\nabla  \; \tx{While $k \leq k_{\mrm max}$, do the following}$:\\
$ \indent \; \diamond  \;  \tx{Use a bisection search method to determine a number $\ze^* > 0$ as large as}$\\
$ \qu \qu \qu \tx{possible for $\ze$ such that the value of $R$ associated with $a, \; b$ and $\ze^*$ is no}$\\
$ \qu \qu \qu \tx{less than $1$. Let $A^*, \; B^*$ and $Q^*$ respectively denote the corresponding}$\\
$ \qu \qu \qu \tx{values of $A, \; B$ and $Q$}$.\\
$ \indent \; \diamond  \;  \tx{If $Q^* < \wh{Q}$, then let $\wh{a} \leftarrow \ze^* a, \; \;
\wh{b} \leftarrow \ze^* b$ and $\wh{Q} \leftarrow Q^*$.  If $A^* = Q^*$, then let}$\\
$ \qu \qu \qu \tx{$a \leftarrow \ze^* a ( 1 + \f{ Q^* - 1 }{5} )$.  If $B^* = Q^*$, then let $b \leftarrow \ze^* b ( 1 + \f{ Q^* - 1 }{5} )$.
Let $k \leftarrow k +
1$}$.\\
$ \nabla  \; \tx{Return $\ze = 1$ as the desired risk tuning parameter and $\wh{a}, \wh{b}$ as}$\\
$ \;  \; \; \; \tx{the weighting coefficients.}$
\\ \hline
\end{tabular}

\vspace{0.05in}

 The intuition behind this algorithm is that $\Pr \{ \tx{Reject} \; \mscr{H}_0
\mid p_0 \}$ and $\Pr \{ \tx{Reject} \; \mscr{H}_1 \mid p_1 \}$ are ``roughly'' increasing with respect to $a$ and $b$, respectively, when the
risk tuning parameter $\ze$ is fixed.

In the execution of the algorithm, we need to compute the probabilistic terms like $\Pr \{ \tx{Reject} \; \mscr{H}_0 \mid p_0 \}$ and $\Pr \{
\tx{Reject} \; \mscr{H}_1 \mid p_1 \}$.  These quantities can be computed by the path counting method of \cite{frazen} or the recursive
algorithm of \cite{sch}.

\subsection{Maximum Number of Observations}

One salient feature of the above algorithm is that the maximum number of observations is absolutely bounded.   Moreover, the maximum number is
the least integer no less than $m$ which satisfies the following equations: \bee  &  & \li (  \f{1 - p_0}{1 - z} \ri )^{1 - z} \li ( \f{p_0}{z}
\ri )^z = \li ( \ze a \ri
)^{\f{1}{m}}, \\
&  &  \li (  \f{1 - p_1}{1 - z} \ri )^{1 - z} \li ( \f{p_1}{z} \ri )^z  = \li ( \ze b \ri )^{\f{1}{m}}, \eee where $z \in (p_0, p_1)$.  To solve
the above equations for $m$, we first eliminate $m$ and obtain
\[
\f{ \ln \li [ \li (  \f{1 - p_0}{1 - z} \ri )^{1 - z} \li ( \f{p_0}{z} \ri )^z \ri ]  }
 { \ln \li [ \li (  \f{1 -
p_1}{1 - z} \ri )^{1 - z} \li ( \f{p_1}{z} \ri )^z  \ri ] } = \f{\ln (\ze a) } { \ln (\ze b) },
\]
from which we find the root $z = z^*$  by a bisection search method. Afterward, we substitute $z = z^*$ into the first equation to obtain the
corresponding $m = m^*$.  Then, the maximum number of observations is equal to $\lf m^* \rf + 1$.  It should be noted that, in the special case
of $a = b$, we have {\small \[ (1 - z) \ln \f{ 1 - p_0 } { 1 - p_1 } + z \ln \f{p_0}{p_1} = 0, \]}  from which we obtain {\small \[ z^* = \f{
\ln \f{ 1 - p_0 } { 1 - p_1 }  } { \ln \f{ (1 - p_0) p_1 } { (1 - p_1) p_0 }   } \] }  and a closed-formed formula for $m^*$.

\subsection{Comparison with TRWA}

\begin{figure}[here]
\centering
\includegraphics[height=8cm]{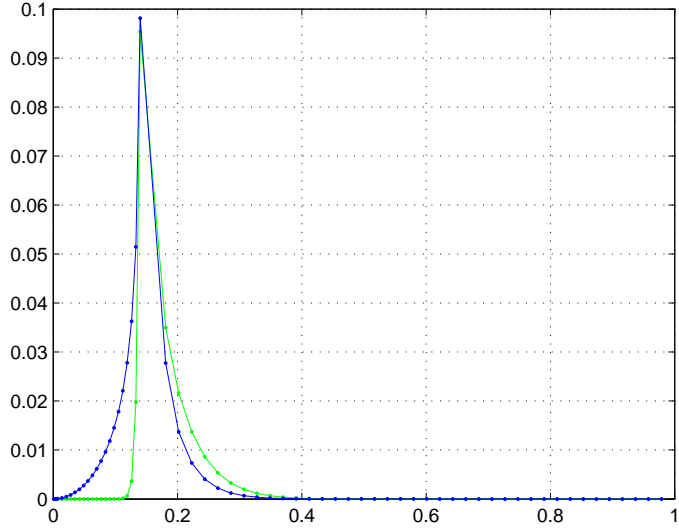}
\caption {Comparison of risks}
\label{fig_Risk}       
\end{figure}

\begin{figure}[here]
\centering
\includegraphics[height=8cm]{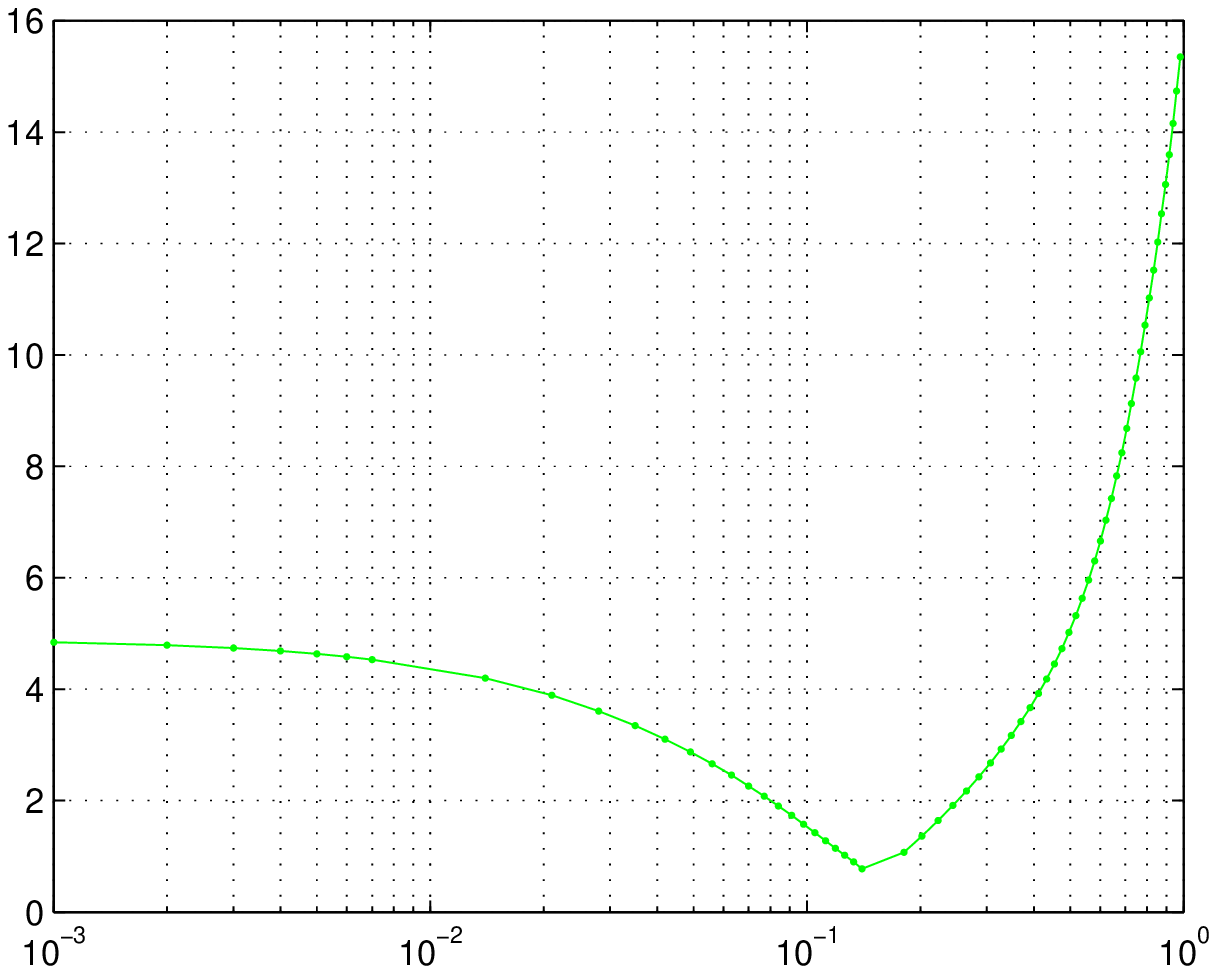}
\caption {Comparison of average number of observations}
\label{fig_Ratio}       
\end{figure}

We have conducted numerical experiments for comparing our detection scheme with TRWA.  For the case of $p_0 = 0.1, \; p_1 = 0.15$ and $\al = \ba
= 0.1$, the risks of our detection scheme (with $\ze = 0.96, \; a = b = 0.1$) and TRWA are respectively shown by the blue and green plots in
Figure \ref{fig_Risk}.   With the same configuration, the ratio between the average number of observations of our detection algorithm to that of
TRWA is shown in Figure \ref{fig_Ratio}. Our computation shows that the new detection algorithm requires a much smaller number of connection
attempts to detect a scanner as compared to TRWA.

\section{Multi-Valued Decision}

\begin{figure}[here]
\centering
\includegraphics[height=8cm]{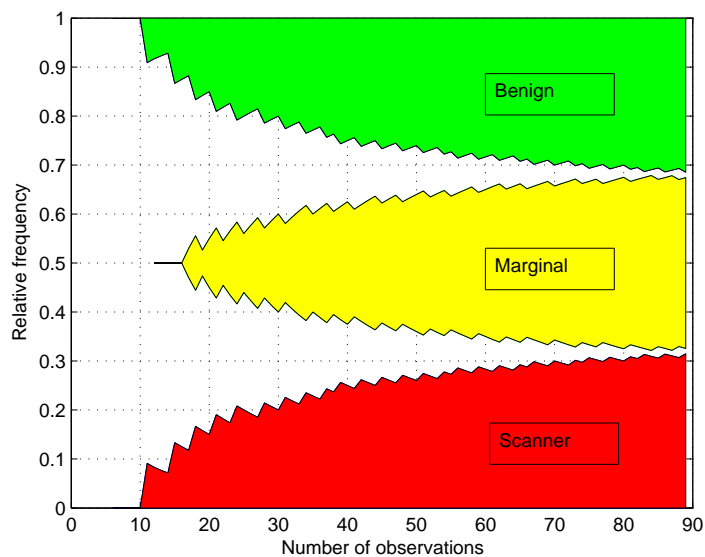}
\caption {Triple Hypothesis Testing}
\label{fig_multidecision}       
\end{figure}

As can be seen from the risk requirement (\ref{riskspec}), there is no specification imposed for users with success rate $p \in (p_0, p_1)$.
This implies that those users can be arbitrarily classified as either scanners or benign users.  In applications, $p_0$ is usually chosen as a
number close to $0$, while $p_1$ is chosen as a number close to $1$.  Therefore, there exists a wide gap between $p_0$ and $p_1$. This indicates
that there is a large portion of ``marginal'' users being cast into either the category of scanners or benign users.   In view of this
situation, we propose to classify the users as three categories: scanner, marginal, and benign. Specifically, let $p_0$ and $p_1$ be two
threshold values such that $0 < p_0 < p_1 < 1$.  We propose to test the following three hypotheses:
\[
\mscr{H}_0: p \leq p_0, \qqu \mscr{H}_1: p_0 < p < p_1, \qqu \mscr{H}_2: p \geq p_1
\]
where hypotheses $\mscr{H}_0, \; \mscr{H}_1$ and $\mscr{H}_2$ corresponds to the categories of ``scanner'', ``marginal'', and ``benign''. Based
on the classification, different actions are taken for the corresponding categories.  To control the probabilities of making wrong decisions, we
impose the following requirement: \bee &  & \Pr \{  \tx{Reject} \; \mscr{H}_0 \mid p  \} \leq \de_0 \qu \tx{for} \; p \in (0,
p_0^\prime],\\
 & & \Pr \{ \tx{Reject} \; \mscr{H}_1 \mid p  \} \leq \de_1 \qu \tx{for} \; p \in [p_0^{\prime \prime}, p_1^\prime],\\
&  & \Pr \{  \tx{Reject} \; \mscr{H}_2 \mid p  \} \leq \de_2 \qu \tx{for} \; p \in [p_1^{\prime \prime}, 1) \eee where $0 < p_0^\prime < p_0 <
p_0^{\prime \prime} < p_1^\prime < p_1 < p_1^{\prime \prime} < 1$ and $\de_i \in (0, 1)$ for $i = 1, 2, 3$. The intervals $(p_0^\prime,
p_0^{\prime \prime})$ and $(p_1^\prime, p_1^{\prime \prime})$ are called indifference zones, since no specification is imposed for controlling
the probability of making wrong decisions for $p$ contained in these intervals.   This problem is actually a special case of the general problem
of testing multiple hypotheses, which has been systematically addressed in our recent paper \cite{Chen1}.  The techniques in \cite{Chen1} offer
a complete solution to the present problem of testing triple hypotheses on the success rate $p$.  As an illustration, assume that \[ \de_0 =
\de_1 = \de_2 = 0.1,
\]
\[ p_0 = \f{1}{3}, \qqu  p_1 = \f{2}{3},
\]
and
\[
p_0^\prime = p_0 - \f{1}{9}, \qu p_0^{\prime \prime} = p_0 + \f{1}{9}, \qu p_1^\prime = p_1 - \f{1}{9}, \qu p_1^{\prime \prime} = p_1 +
\f{1}{9}.
\]
By virtue of the technique of \cite{Chen1}, we have obtained a sequential testing scheme shown by Figure \ref{fig_multidecision}, where the
bottom, middle and upper shaded areas represent the acceptance regions of $\mscr{H}_0, \mscr{H}_1$ and $\mscr{H}_2$, respectively. The stopping
and decision rules can be stated as follows:

 If the sample path, which can be represented by the plot of the relative frequency $\wh{p}_n$ versus the number $n$ of observations, hits a
 shaded region, then terminate the observational process.
 At the termination of the observational process, accept the hypothesis of which the acceptance region is hit by the sample path.

\begin{figure}[here]
\centering
\includegraphics[height=8cm]{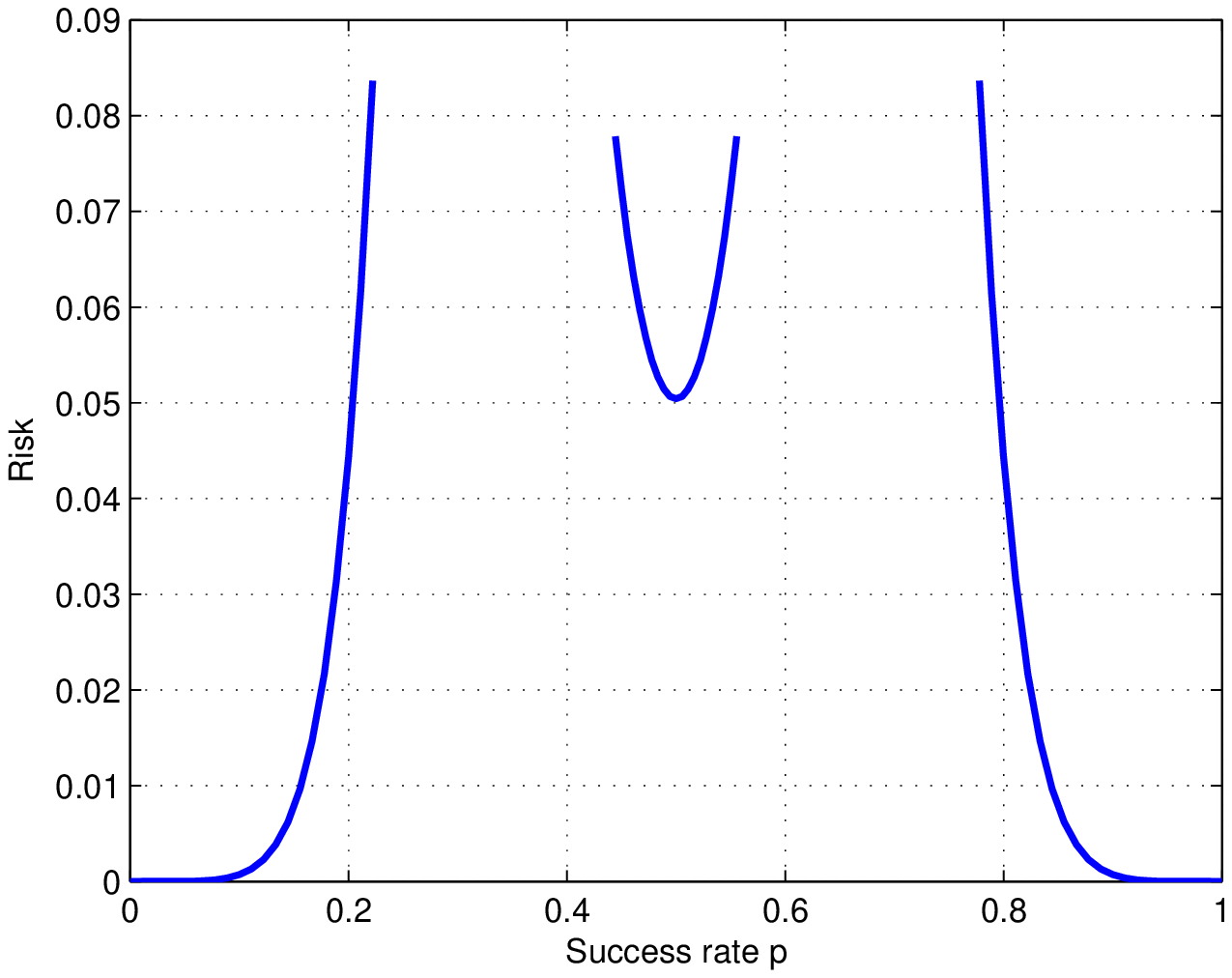}
\caption {Risk}
\label{fig_Riskm}       
\end{figure}

In Figure \ref{fig_Riskm},  we plot the risk, $\Pr \{ \tx{The decision is incorrect} \mid p  \}$, versus the success rate $p$.  It can be seen
 that the risk requirement is satisfied for any $p \in (0, 1)$ not contained in the indifference zones.

\section{Conclusion}

We have developed new sequential methods for detecting portscanners.  In addition to guaranteeing the risk requirement, our algorithm is
efficient when the success rate assumes values other than the threshold values.  Moreover, the required number of observations is absolutely
bounded. Furthermore, we have proposed a framework of multi-valued decision for testing portscanners.

\end{document}